\documentclass[%
 preprint,
 amsmath,amssymb,
 aps,
]{revtex4-2}

\usepackage{graphicx}
\usepackage{dcolumn}
\usepackage{bm}
\usepackage{xcolor}

\begin{document}

\preprint{ }

\title{Deep learning-assisted imaging through stationary scattering media}

\author{Siddharth Rawat}%
\author{Jonathan Wendoloski}
\author{Anna Wang}
 \email{anna.wang@unsw.edu.au}
 \affiliation{School of Chemistry, UNSW Sydney, NSW 2052 Australia}


\begin{abstract}
Imaging through scattering media is a challenging problem owing to speckle decorrelations from perturbations in the media itself. \textcolor{black}{For in-line imaging modalities, which are appealing because they are compact, require no moving parts, and are robust, negating the effects of such scattering becomes particularly challenging. Here we explore the effect of stationary scattering media on light scattering in in-line geometries, including digital holographic microscopy. We consider various object-scatterer scenarios where the object is distorted or obscured by additional stationary scatterers}, and use an advanced deep learning (DL) generative methodology, generative adversarial networks (GANs), to mitigate the effects of the additional scatterers. Using light scattering simulations and experiments on objects of interest with and without additional scatterers, we find that conditional GANs can be quickly trained with minuscule datasets and can also efficiently learn the one-to-one statistical mapping between the cross-domain input-output image pairs. Training such a network yields a standalone model, that can be used later to inverse or negate the effect of scattering, yielding clear object reconstructions for object retrieval and downstream processing. Moreover, it is well-known that the coherent point spread function (c-PSF) of a stationary scattering optical system is a speckle pattern which is spatially shift variant. We show that with rapid training using only 20 image pairs, it is possible to negate this undesired scattering to accurately localize diffraction-limited impulses with high spatial accuracy, therefore transforming the earlier shift variant system to a linear shift invariant (LSI) system.
\end{abstract}

\maketitle

\section{INTRODUCTION}

Imaging through scattering media has recently gained attention due to a myriad of desirable applications such as deep tissue imaging\cite{izatt1994optical,denk1990two}, imaging through fog\cite{satat2018towards, leith1992imaging, kijima2021time}, underwater imaging \cite{kocak2008focus, hou2009simple} and more. Laser speckles are a nuisance when imaging through disordered media such as a diffuser or biological tissue. While methods such as cleaning and slicing can be used during sample preparation to minimize the adverse scattering from unwanted scatterers, these methods are invasive and can physically damage the sample~\cite{tian2021tissue, costa2019optical}. It is thus desirable to invert or negate the effect of this extra scattering by non-invasive or computational means. 

The main bottleneck for the analytical approach lies in accurately modelling and characterizing the whole scattering process. This characterisation becomes challenging owing to the sheer number of variables involved and the associated degrees of freedom present within the scatterer and scattering media. 

Nonetheless, optical scattering still can be regarded as a deterministic process\cite{li2018deep} and a series of strategies have been developed to negate the effect of scattering. Iterative wavefront shaping\cite{kubby2019wavefront, horstmeyer2015guidestar}, the transmission matrix\cite{boniface2020non, popoff2010measuring, ma2014time} approach, speckle correlation\cite{katz2014non, ruan2020fluorescence}, and use of guide stars for point spread function (PSF) estimation~\cite{edrei2016memory,bertolotti2015unravelling, schneider2018guide}, are all useful but come with their own limitations. In the iterative wavefront shaping method, the involvement of localized reporters and the number of trials involved to get the final optimized wavefront can both present challenges. The transmission matrix (TM) approach requires a spatial light modulator hence complicated opto-acoustic configurations and has limited depth of field\cite{boniface2020non}. Speckle correlation methods are less invasive than the aforementioned two methods: the object field is reconstructed from the autocorrelation of the field recorded at the sensor, using Fienup-type iterative phase retrieval algorithms. The only drawback of this approach is the limited angular range due to the memory effect\cite{katz2014non}. PSF estimation based on a guide star and blind deconvolution using algorithms such as Richardson-Lucy is also limited by the small angular range of reconstruction\cite{edrei2016memory, bertolotti2015unravelling, schneider2018guide}.

In this manuscript we treat the scattering as an image-to-image translation problem in order to use recent advancements in machine and deep learning. Such data-driven approaches have gained traction in solving a range of inverse problems in optics including phase unwrapping\cite{rawat2021accurate}, image retrieval\cite{yan2020deep}, defect detection\cite{chien2019complex}, and coherent imaging\cite{rivenson2019deep,zhang2021neural, zhang2018motility, yoon2020deep}. Li and coworkers\cite{li2018deep} proposed a highly scalable new deep learning framework that encapsulates a range of statistical variations for the trained model for robust reconstructions, despite speckle decorrelations. The convolutional neural network (CNN) architecture used was fairly simple, being essentially an autoencoder. We expect more sophisticated architectures to yield reconstructions that are qualitatively and quantitatively more similar to the ground truth. \textcolor{black}{For example recently, Lai and coworkers used a YGAN architecture trained on thousands of image pairs to retrieve pairs of objects from speckle patterns~\cite{lai_reconstructing_2021}.}

\textcolor{black}{Here, we use conditional generative adversarial networks or cGANs\cite{isola2017image} to analyze images from in-line coherent imaging setups.} We show that cGANs can be used to efficiently learn the one-to-one statistical mapping between the scattered field (diffused/speckle patterns) and the corresponding clear object representations with as few as 20 image pairs. With rapid training, the signal from the object of interest can be retrieved from a highly-scattering system. Importantly, the recovered signal is faithful enough to the original image for further analysis, leading to excellent quantitative feature extraction \textcolor{black}{of moving objects in in-line imaging setups. As a result, objects or impulses can be localized to great accuracy, and cGANs can be used to preserve the 3D-precision tracking ability of holographic microscopy even in the presence of unwanted scatterers.}

\section{NETWORK ARCHITECTURE}

\begin{figure}[htbp]
\centering
\fbox{\includegraphics[width=\textwidth]{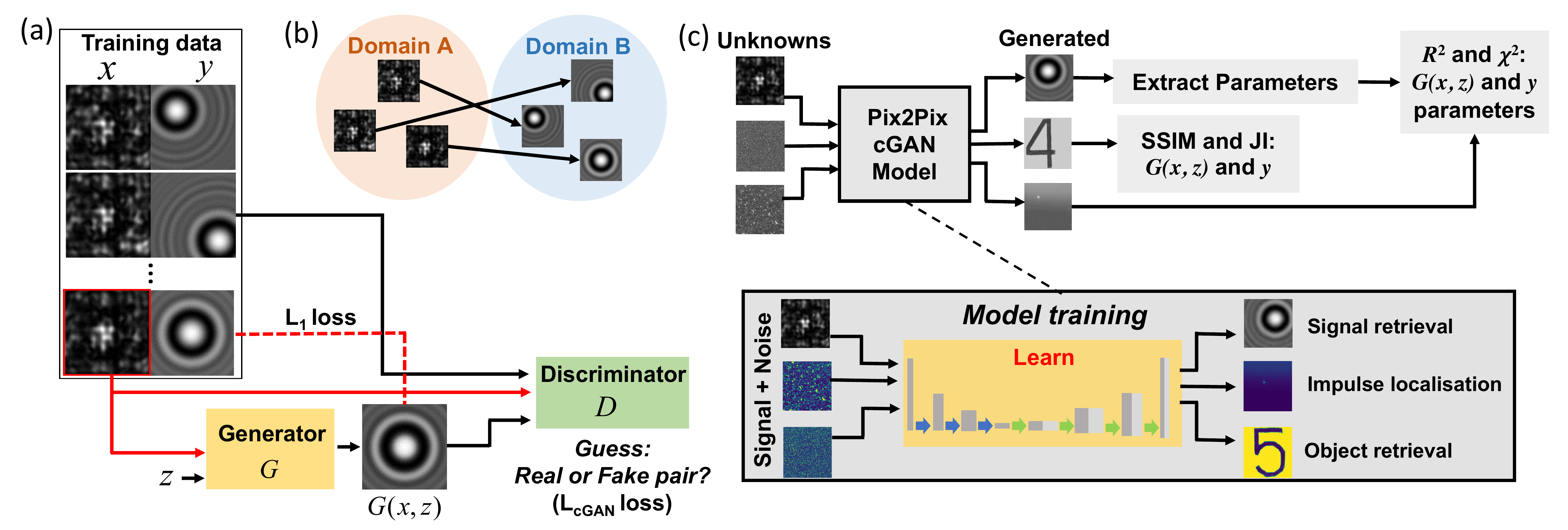}}
\caption{\textbf{Principle and working of a cGAN network architecture for several image translation tasks.} (a) Complete cGAN architecture: Paired training dataset is used to train the network consisting of a generator $G$ and a discriminator $D$ block.  Generator outputs a synthetic translation for a given input $x$ and a random noise vector $z$. The discriminator $D$ is a tasked to classify between real and fake image pairs. Alongside the L$_\mathrm{cGAN}$ loss, $L_1$ loss is also considered, so that the synthetically generated images remains closer to the ground truth. (b) As a result, data is paired between two image Domains A and B. (c) Simulation workflow from training to inference/testing: After training, a standalone model can then be used to perform image translation tasks to generate $G(x,z)$ (Domain B) from non-discernible images $x$ (Domain A), resulting in signal retrieval, impulse localization, or object retrieval.}
\label{fig1}
\end{figure}

The principle and workflow of a cGAN network architecture for several image translation tasks we employ is illustrated in Fig.~\ref{fig1}. Fig.~\ref{fig1}(a) shows the complete training of a typical cGAN network architecture; paired image data is used in training the cGAN consisting of a generator and a discriminator block. In this architecture, generator \textit{G} is tasked to produce a synthetic translation of the training data in a conditional sense. The discriminator \textit{D} is tasked to classify whether the inputted image pair is real or fake. Both generator \textit{G} and discriminator \textit{D} compete to optimise their own objectives, and hence termed adversaries. 

Alongside the conditional GAN loss L$_\mathrm{cGAN}$, additional $L_1$ loss is computed to maintain the similarity between the generated and ground truth images. A standalone model is then obtained after the completion of training (see Supplementary Methods in Supplement 1). Fig.~\ref{fig1}(b) shows data pairing between two domains i.e., domain A and domain B. Likewise, any cross-domain data pairs can be formed for training a cGAN network. Fig.~\ref{fig1}(c) shows the complete workflow from training to inference — after model training, a standalone Pix2Pix cGAN model is obtained, that can then be used to translate unknown domain A images to domain B images. Thereafter, several crucial parameters can be extracted from these generated images and error metrics.

\section{PHASE CONJUGATION ROBUSTNESS}
\subsection{Mathematical Formulation of First and Second Order Speckle Intensity Autocorrelation Functions}

Before proceeding with the cGAN, we investigated the speckle field and intensity autocorrelation for a scenario involving scattering from a thin diffuser layer. For further details please see Section 1 in Supplement 1. In brief, we performed second order speckle statistics for the electric field  $A(x^{'})$ at the observation plane and found that the field autocorrelation function $C_{A}(x_1^{'}, x_2^{'}) \equiv \langle A^{\ast}(x_1^{'}) A(x_2^{'}) \rangle $ \cite{vellekoop2008controlling} ($\ast$ denoting the complex conjugate) between any two coordinate points $(x_1^{'}, x_2^{'})$ in the field $A(x^{'})$ is proportional to $|f(x)|^2$. In other words, the field in the observation plane and the field in the object plane are correlated despite randomization. We also showed that first and second order speckle intensity autocorrelation functions are related and are independent of the physical properties of the scattering layer. 

Given this correlation, we then performed scattering simulations based on scalar diffraction theory to determine the robustness of simple phase conjugation (see Supplement 1 and Figures S2-S7). In brief, we used the Angular Spectrum Propagation (ASP) method to simulate forward and backward scattering from an object in presence of a stationary scattering media layer such as a random phase mask (e.g. diffuser). In the simulation, we assumed that the diffuser phase can be estimated using interferometry and can be later used for phase conjugation for object retrieval. 

However, we found that interferometric phase estimation methods are sensitive to noise; translating the diffuser by a  few micrometers renders the earlier phase estimation useless for phase conjugation. This poses a significant challenge to using the interferometric scheme for object retrieval and alternative methods are thus warranted.

In the following sections, we test our cGAN-based approach on systems where the main scatterer of interest is in motion, either from diffusion or from manual translation.

\section{MICROSCOPIC OBJECT RETRIEVAL}
\subsection{Object Retrieval from Behind a Single Scattering Layer}
\label{holo_3D_layer}

Digital holographic microscopy (DHM) is one key imaging technology that uses coherent light and is hence severely affected    by additional scatterers. Though DHM has been used for precision tracking to great success\cite{barkley2019holographic}, dilute systems are typically used   because the interference fringes of neighboring objects limits what information can be recovered about the object of interest.

We test our approach on holograms of a primary scatterer (\textit{P}) positioned near a layer of secondary scatterers (\textit{S}), as shown in  Fig.~\ref{fig2}(a, g). The \textit{S} layers consist of 1-$\mu$m-diameter spheres of refractive index $n_S$ = 1.59, typical of polystyrene latex microspheres. The primary scatterer \textit{P} is 2-$\mu$m-diameter of refractive index $n_P$ = 1.42, approximating that of biological material. The surrounding medium’s refractive index is $n_{med}$ = 1.33, representing aqueous buffers or water. The wavelength of the light source is $\lambda$ = 632.8 nm.

To train the model, we generate image pairs with and without the scattering layer. For \textit{P} obscured by a single \textit{S} layer, we simulated a total of 150 holograms for a mobile \textit{P} and stationary \textit{S}, $x = H_{P+S}$ (see Fig.~\ref{fig2}(b) and Supplementary Methods in Supplement 1). We also simulated corresponding holograms of \textit{P} by itself $y = H_P$. The position coordinates for \textit{P} were chosen to mimic Brownian motion using a Wiener process, a continuous-time stochastic process with stationary independent real increments, trajectory shown in Supplement 1 (Supplementary Methods) and \textbf{Visualization 1}. After training on 100 image pairs $x = H_{P+S}$ and $y = H_P$, we selected a model (Figure S8) to generate \textit{S}-free holograms (see Supplementary Methods in Supplement 1 and \textbf{Visualization 2}) to test against the remaining 50 holograms.

As shown in Fig.~\ref{fig2}(c), the qualitative agreement between cGAN-generated holograms $G(x, z) = H_{cGAN}$ and ground truth holograms $y = H_P$ appeared excellent, so we proceeded to quantitatively verify the performance of our model. We fitted a Lorenz-Mie scattering model (see Supplement 1) to each of the holograms $H_{P+S}$ and $H_{cGAN}$ using the package HoloPy\cite{barkley2019holographic, wang2016tracking, wang2014using}. The fitted results for $H_{P+S}$ and $H_{cGAN}$ were compared to the ground truth values used for simulation (Fig.~\ref{fig2}(d)). While the $x_p$ and $y_p$ coordinates for both $x = H_{P+S}$ and $G(x, z) = H_{cGAN}$ could be recovered, the fit results for the axial position $z_p$ of the scatterer were vastly better for $H_{cGAN}$. The root-mean-square errors RMSE for localizing \textit{P} reveals excellent performance in all three dimensions for $G(x, z) = H_{cGAN}$ (Fig.~\ref{fig2}(e)). 

The high quality of fit for $H_{cGAN}$ corresponds to excellent recovery of the hologram’s qualitative features. The sum of the pixel-by-pixel squared differences between the best-fit hologram and either $x$ or $G(x, z)$, $\chi^2$, and $R^2$ were vastly improved by the cGAN model translation (Fig.~\ref{fig2}(f)). Similarly-promising results were obtained for a scenario where the axial positions of \textit{P} and \textit{S} are reversed, and the axial position of \textit{P} was allowed to change (see Section 2.5 in Supplement 1, \textbf{Visualization 3}, \textbf{Visualization 4}, and Figures S9-10). For complete training and testing parameters see Table S1.

\subsection{Object Retrieval from Between Two Scattering Layers}
\label{holo_3D_layer_1}

\begin{figure}[htbp]
\centering
\includegraphics[width=\textwidth]{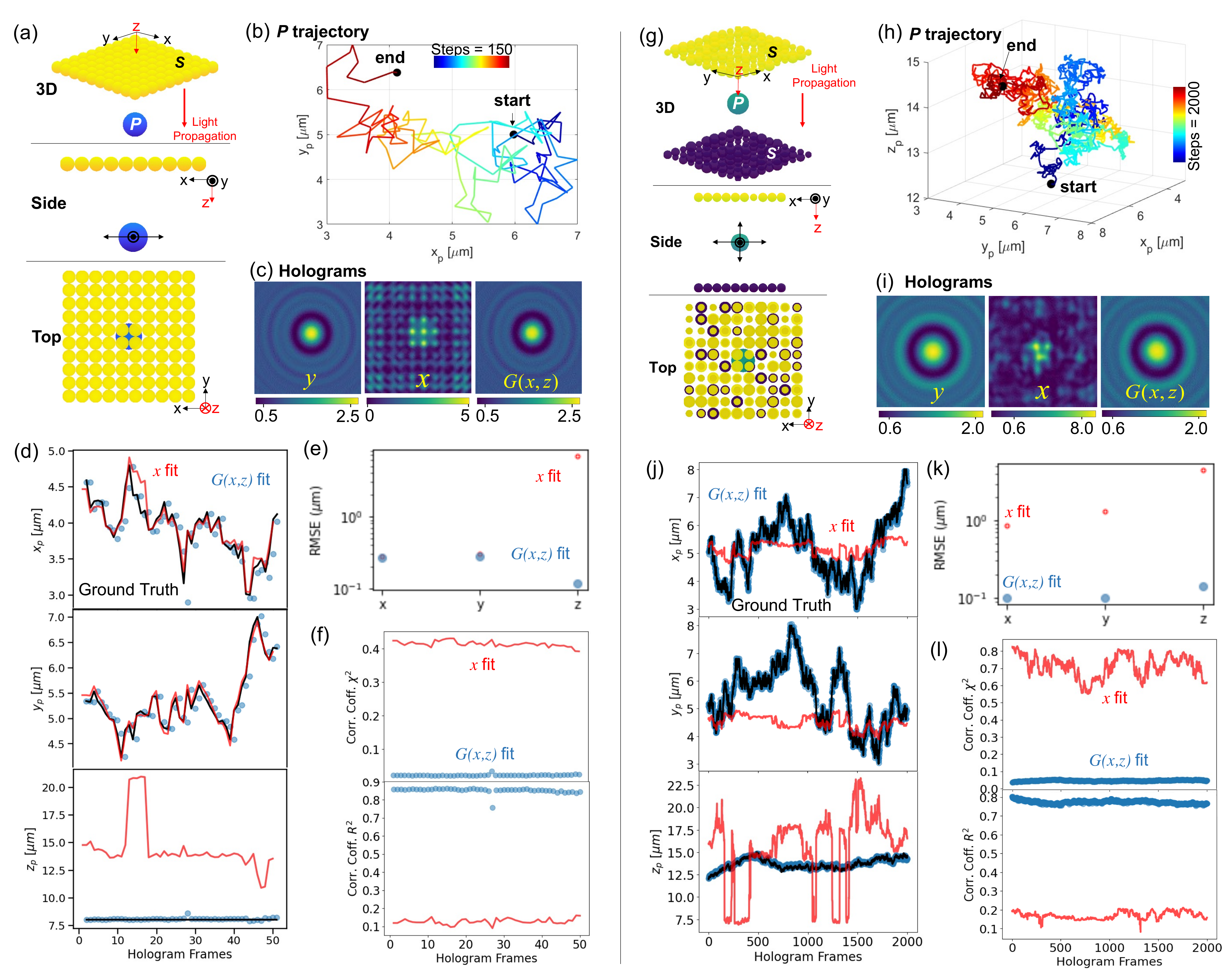}
\caption{Caption next page.} 
\label{fig2}
\end{figure}

Holograms in the presence of a single scattering layer \textit{S} still showed recognizable features of the primary scatterer \textit{P}. We were thus compelled to explore a more adverse scenario: the addition of another scattering layer \textit{S'} renders \textit{P} barely visible (Fig.~\ref{fig2}(g)), and furthermore, we allow $P$ to diffuse in 3D (Fig.~\ref{fig2}(h)). Additionally, the radii $r_i$ of the spheres in \textit{S} and  \textit{S'} were randomly sampled from a discrete uniform distribution $r_i \stackrel{}{\sim} U[.1,\; .5]$.

By allowing $P$ to explore 3D space, we would also need the training set to span 3D space. To mitigate the concomitant increase in training time, we postulated that it may be possible to train the cGAN network on a series holograms of $P$ diffusing with discrete integer steps, despite the testing dataset being for the $P$ diffusing with sub-integer steps. For a system with a single \textit{S} we found that this approach worked effectively (Figures S9-10), so we then applied it to the system with both \textit{S} and \textit{S'}. The training dataset consisted of a total of 144 simulated hologram image pairs consisting of $x = H_\mathrm{P+S+S'}$ and $y = H_\mathrm{P}$ (Fig.~\ref{fig2}(i)). 

We generated the testing dataset separately using a 3D random walk with sub-integer step sizes (see Fig.~\ref{fig2}(h) and supplementary \textbf{Visualization 5}). Even though the network was trained on discrete data (Figure S11), it accurately inferred the real testing dataset with particle localization in all three dimensions being very close to the ground truth (see Fig.~\ref{fig2}(j)). Notably, the RMSE for particle localization was less than 0.2 $\mu$m in all three dimensions (Fig.~\ref{fig2}(k)), whereas the particle could not be localized with the fitting routine for unprocessed holograms $x = H_\mathrm{P+S+S'}$. 

For the dual scattering layer case, the overall quality of fit for $H_\mathrm{cGAN}$ was excellent (Fig.~\ref{fig2}(l)), just as it was for the single scattering layer case. The $\chi^2$ and $\textit{R}^2$ values were vastly improved by the cGAN model translation (see Figure S11 and \textbf{Visualization 6}). For complete training and testing parameters see Table S1.

\clearpage

\clearpage 

\textbf{Caption for Figure~\ref{fig2}: Spatial parameter extraction by fitting a Lorenz-Mie scattering model to the cGAN-generated and noisy holograms for a mobile primary scatterer ($P$) in two different scattering scenarios.} (a) A primary scatterer $P$ diffuses freely in two dimensions in front of a stationary scattering layer $S$. (b) The complete path of P (150 steps) within bounds i.e. 3 $\leq$ ($x_p$, $y_p$) $\leq$ 7 and $z_p$ = 8. (c) The ground truth $y$ = $H_\mathrm{P}$, cGAN-generated $G(x,z)$ = $H_\mathrm{cGAN}$, and the complex hologram $x$ = $H_\mathrm{P+S}$. (d) Fit results for localizing $P$ in $x = H_\mathrm{P+S}$ and $G(x,z)$ = $H_\mathrm{cGAN}$, versus ground truth values. (e) The root mean squared error (RMSE) for localizing $P$ in the axial direction is an order of magnitude improved using our cGAN model. (f) Error metrics $\chi ^2$ and $R^2$ reveal excellent recovery of hologram features by the trained model. (g) A primary scatterer $P$ diffuses freely in three dimensions between two stationary scattering layers $S$ and $S'$. (h) The complete 3D path of $P$ (2000 steps) within bounds i.e. 3 $\leq$ ($x_p$, $y_p$) $\leq$ 8 and 12 $\leq$ $z_p$ $\leq$ 15. (i) The ground truth $y$ = $H_\mathrm{P}$, cGAN-generated $G(x,z)$ = $H_\mathrm{cGAN}$, and the complex hologram $x$ = $H_\mathrm{P+S+S'}$. (j) Fit results for localizing $P$ in $x = H_\mathrm{P+S+S'}$ and $G(x,z)$ = $H_\mathrm{cGAN}$, versus ground truth values. (k) The root mean squared error (RMSE) for localizing $P$ in all directions is an order of magnitude improved using our cGAN model. (f) Error metrics $\chi^2$ and $R^2$ shows excellent recovery of hologram features by the trained model. For further details please see Visualizations 1-6.

\section{EXPERIMENTAL OBJECT LOCALIZATION AND RETRIEVAL}
\subsection{Accurate Object Localization in a Spatially Shift Variant System}
\label{nonlinear}
Given the promising results thus far, we sought to experimentally verify whether the same approach could be used to negate effects of a scattering layer on a point spread function (PSF). Ordinarily, imaging a point source such as a sub-micron fluorescent bead or a diffraction limited point source can be used to determine the PSF of an incoherent system or a coherent point spread function (c-PSF) of coherent imaging system \cite{schneider2018guide}. For an ideal optical system with very high space-bandwidth product, the PSF is linear and shift invariant. However, these two assumptions do not hold for imaging systems with aberrations imposed by imperfections owing to undesirable scattering. As an extreme example, imaging point sources behind a stationary scattering layer can lead to formation of complex patterns such as laser speckles when using coherent light sources. This scattering makes the c-PSF (a speckle pattern) non-linear and spatially variant, which further limits the imaging field of view owing to memory effects \cite{katz2014non,schneider2018guide}. 

\begin{figure}[htbp]
\centering
\fbox{\includegraphics[width=\textwidth]{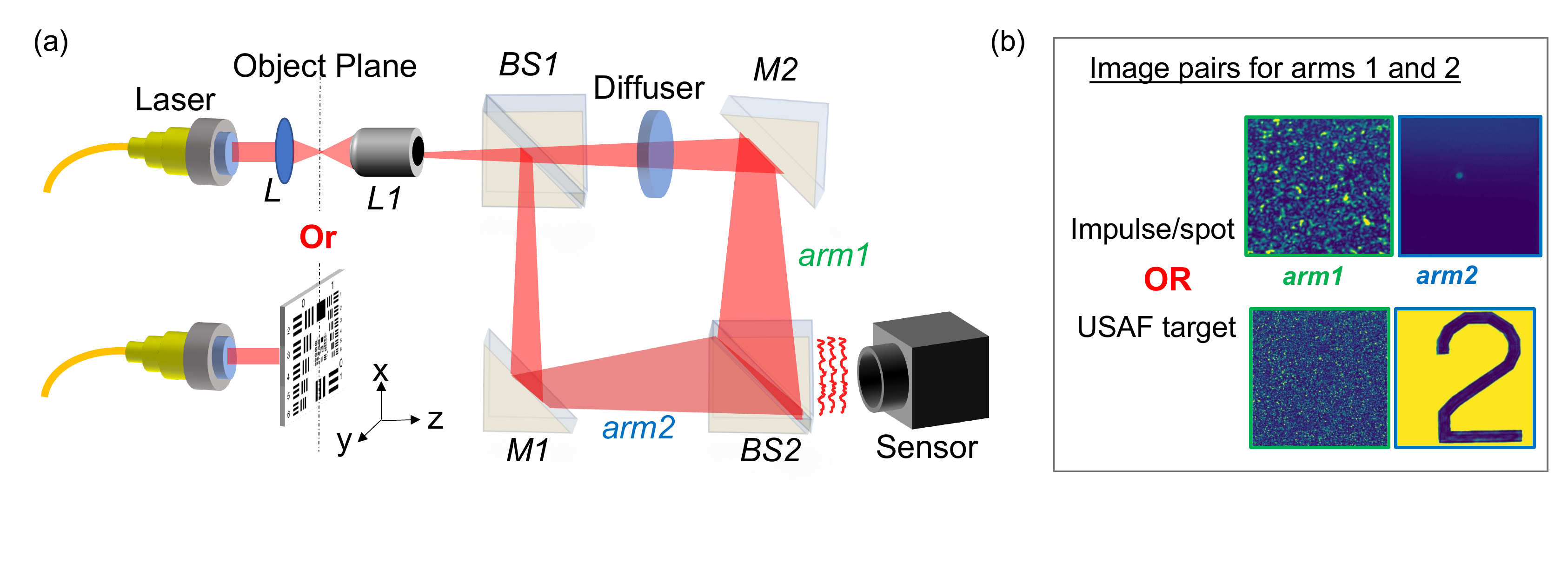}}
\caption{\textbf{Experimental setup for data acquisition of two different input types in the object plane.} (a) A low NA lens \textit{{L}} produces a diffraction limited spot in the image plane. Likewise, a USAF test target can be placed in the object plane. An objective lens (4x) \textit{{L1}} is then used to image the object plane. Beam splitter \textit{{BS1}} splits the amplitude of the diverging spherical beam into two arms \textit{{arm1}} and \textit{{arm2}}. Mirrors \textit{{M1}} and \textit{{M2}} direct the beam to another beamsplitter \textit{{BS2}}. A sensor is then placed after \textit{{BS2}} to record the image pairs. (b) By sequentially blocking either \textit{{arm1}} or \textit{{arm2}}, image pairs $x$ and $y$ can be recorded for both the impulse/spot and the USAF target.}
\label{fig3}
\end{figure}

We used a Mach Zehnder setup as shown in Fig.~\ref{fig3}(a) to acquire c-PSFs of a point source with and without a scattering layer. Unlike a typical Mach Zehnder setup, which is used for interference between two beams, our use of the setup was to ensure stability of the two arms during imaging. Specifically, a diffraction limited spot was generated by a short focal length lens (Thorlabs, C171TMD-B), placed after the fiber collimated laser source (Thorlabs, PAF2-A7B), then imaged by a low NA lens (Nikon Plan Fluor, 4X/0.13). The amplitude of the optical field was split using a beam splitter (\textit{BS1}). A ground glass diffuser (Edmund Optics, 220 grit) was placed in between \textit{BS1} (Thorlabs, CCM1-BS013) and a mirror \textit{M2} (Thorlabs, BB03-E02) to produce a speckled PSF which is non-linear and shift variant; this constitutes the first arm, \textit{arm1}. The beam in the alternate arm \textit{arm2} was a clear and magnified image of the point source or impulse. The main advantage of this geometry is that it keeps the optical path lengths in the two arms matched, allowing us to record (Basler, acA1920-155um) the speckled image from \textit{arm1} and clear image of the object from \textit{arm2} simply by alternately blocking the arms (Fig.~\ref{fig3}(b)). For further discussion of the propagation, see Supplement 1.

Twenty-five translated impulses along were collected, as shown in Fig.~\ref{fig4}(a) along with the rasterization path (red dotted line). For every impulse, a speckle pattern was also recorded (see Fig.~\ref{fig3}(b), top, and Fig.~\ref{fig4}(b)). Thereafter, the paired dataset was split into testing and training sets (see Fig.~\ref{fig4}(a)). Training this minuscule paired dataset (orange in Fig.~\ref{fig4}(a)) only took $\sim$ 10 minutes in a Google Colaboratory environment (Figure S12). The five measured/test speckle patterns (see white bounding box in Fig.~\ref{fig4}(a)) which were not used in training were then inputted individually into a standalone model model acquired after training. The model generated reconstructed images, a composite of which are shown in Fig.~\ref{fig4}(b). 

By comparing the centroids of the original and cGAN-generated impulses, we confirmed that the impulse recovery was excellent not only qualitatively, but also quantitatively. The root mean square error (RMSE) between the original and cGAN-generated spot coordinates in $X$ is 1.0864, and $Y$ is 1.0687 pixels (see Fig.~\ref{fig4}(c)) compared to a spot diameter of 15 pixels (FWHM 10 pixels). We therefore have shown that is possible to localize diffraction-limited objects in a spatially shift-variant optical system with high spatial accuracy, despite the c-PSF presenting as random noise. Complete training and testing parameters are tabulated in Table S1.

\subsection{Imaging Through Stationary Scattering Media}
\label{imaging_scat_media}
\begin{figure}[htbp]
\centering
\fbox{\includegraphics[width=\textwidth]{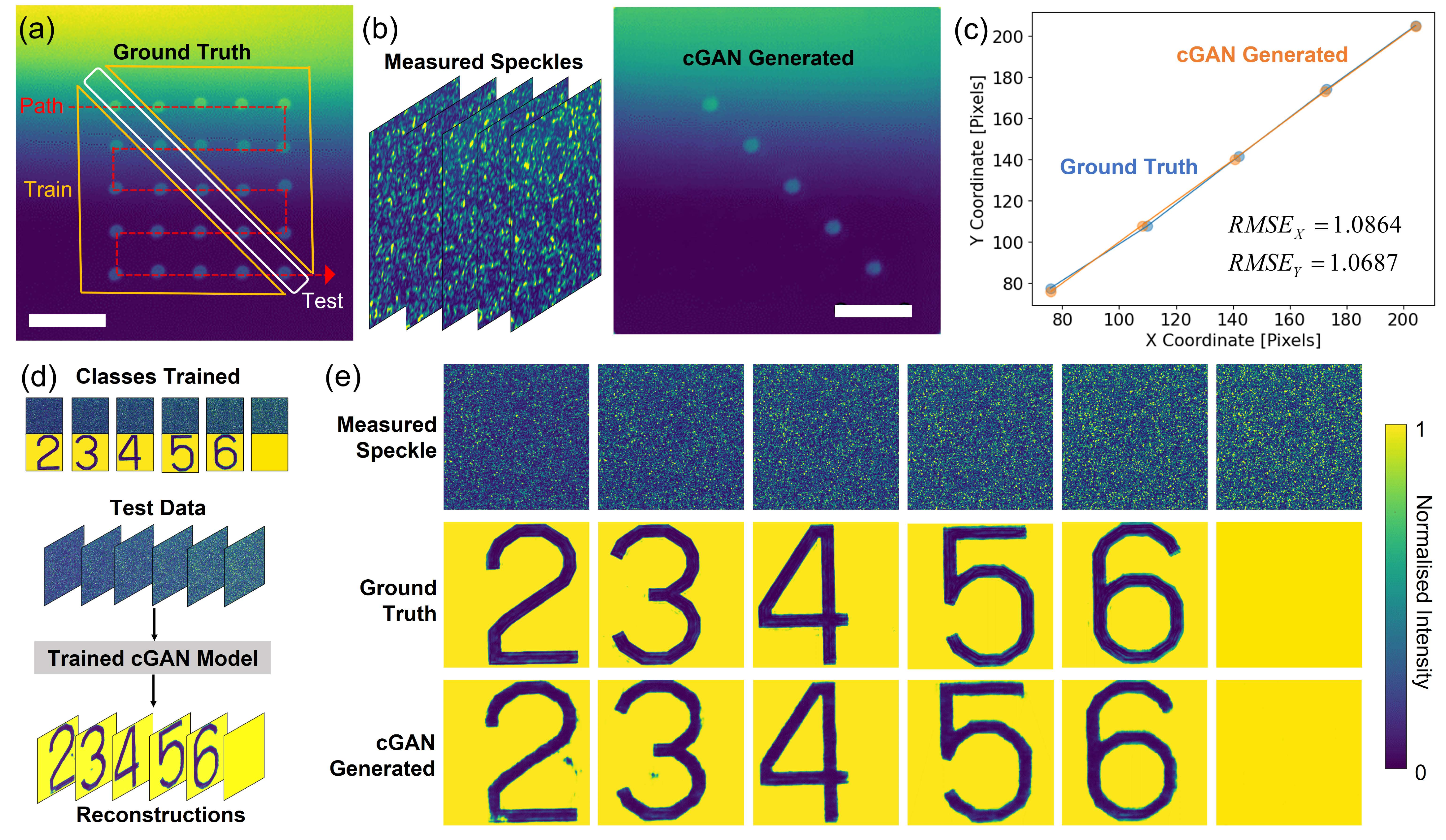}}
\caption{ \textbf{Object retrieval and localization from unknown speckle patterns.} (a) A composite image of the 25 translated impulses. The impulses within orange bounding boxes were used for training the cGAN network and the diagonal impulses within the white bounding box were used for testing. The red dotted line shows the rasterisation path while recording these impulses. (b) The 5 test speckle patterns for the translated impulses in the object plane (c.f. white bounding box in (a)), and their corresponding cGAN-generated images $G(x,z)$. (c) The generated images $G(x,z)$ were compared with the ground truth images $y$. Low RMSE in $X$ and $Y$ coordinates show excellent recovery of the impulses by the model. All the individual images are 256 x 256 pixels and the white scale bar represents 50 pixels. (d) For the USAF target object recovery, six class types were trained along with their speckle images. (e) Comparison between the cGAN-generated $G(x,z)$ and ground truth test images $y$ demonstrate excellent object recovery from the speckle patterns $x$.}
\label{fig4}
\end{figure}

Having established that our trained model can localize point sources with great accuracy, we then attempted to image objects through a scattering layer. The specimen chosen was a USAF target (Edmund Optics, 36-275), placed into the object plane of the objective lens \textit{L1} in place of the lens from the previous experiment. To collect data, we translated the USAF structures of interest (group 4) in the objective's field of view, then alternately blocked \textit{arm1} to collect the clean object images, or \textit{arm2} to record the corresponding speckle image (see Fig.~\ref{fig3}(b), bottom). A total of six classes were trained (Fig.~\ref{fig4}(d), top), with the whole dataset consisting of 90 images pairs or 15 image pairs per class for six classes. 

Importantly, each image pair had a slightly different position of the USAF target relative to the diffuser. Two-thirds of the data per class was used for training (see Fig.~\ref{fig4}(d), Supplement 1, and Figure S13), the remaining one-third was used for testing. 

Despite the testing set having a diffuser position on which the model was not trained, we found that the object recovery was excellent. Fig.~\ref{fig4}(e) compares ground truth test images and generated images, when the recorded test speckle pattern was inputted into the best model. The average time to reconstruct from a speckle pattern of 256 x 256 pixels was $\sim$ 1.43 seconds. Complete training and testing parameters are tabulated in Table S1. We believe this is the first experimental demonstration of signal retrieval of a translated object from behind a diffuser without using quasi-planar devices such as spatial light modulators.

\section{DISCUSSION AND CONCLUSIONS}

Our results show that cGAN-based-models can be efficiently trained to generate faithful representations $G(x,z) \sim y$ (Figure S1). The size of the training set did not exceed 144 image pairs for this work (see Table S1 for summary of model details). Nonetheless, the object reconstructions were not only qualitatively comparable to the ground truth, but were also accurate enough to enable quantitative post-processing.

The first simulation case described a scenario in which a primary scatterer (\textit{P}) was diffusing freely near a single stationary scattering layer, causing attenuation in the incident beam. Only 100 images pairs were used for training in this case. Surprisingly, the extracted parameters based on this minuscule training set were close to the ground truth values with very low RMSE. Likewise, in second simulation case (see Figure S10) we simulated \textit{P} diffusing in three dimensions near scattering layer \textit{S} but with its position reversed with respect to the previous case. For this scenario we chose a slightly larger training set consisting of $144$ image pairs. The rationale is that due to added degree of freedom in the axial direction, the diffuse interference pattern changes in addition to being translated. This prompted us to incorporate more image pairs for various 3D spatial locations in the training step. Just like the previous case, the RMSE between the extracted parameters and the ground truth values was low, suggesting that the network learned efficiently despite the small training set.

In the last simulation, we demonstrated that a network can be efficiently trained on little data, even when the object of interest is obscured by two scattering layers \textit{S} and \textit{S'}. Our training image pairs were formed by simulating discrete positions of \textit{P} with integer step size, whereas the testing dataset consisted of image pairs for the motion of \textit{P} with real step sizes. The testing dataset fitted very well despite the network training on integer steps, demonstrating that with cGANs it is possible to have accurate object reconstructions with small training datasets. This strategy is computationally less expensive as it lessens need to train with a large dataset, whilst retaining high reconstruction accuracy and low RMSE of extracted parameters.

We then employed a modified MZ interferometer (Fig.~\ref{fig3}(a)) for experimental validation of our proposed method for imaging through stationary scattering media. We showed that our proposed DL framework employing a cGAN architecture can be used to transform a spatially shift variant optical system to a shift invariant system for accurate object localization and that it is possible to localize diffraction limited objects such as impulses in a diffusive media (Fig.~\ref{fig4}(a-c)). The implication of these results is that this concept can be expanded to image any class of objects with pixel-level accuracy. Indeed we found that even when translating the object of interest, a USAF target, to an untrained location, the model could still generate faithful object retrievals.

The robustness of cGANs with diffused and speckled data warrants further discussion. The cGAN network architecture employs convolutional neural networks in the generator and discriminator blocks. Based on our results we hypothesize that cGANs are efficient at learning the input-output mapping, even when the input pixel representation is randomized. This randomized input is further down-sampled to a higher dimensional latent space (non-discernible by humans) in the generator. We found that in contrast to other methods, there is no strict requirement for a direct visual correspondence between the input and output image pairs while training these networks. This condition is well suited for purposes where the objects of interest may be in motion or translated. To further add to this proposition, Kim \textit{et al.}~\cite{kim2017learning} showed that GANs such as DiscoGANs can be trained to learn cross domain relations efficiently. Similarly Wolf \textit{et al.} \cite{taigman2016unsupervised} showed that it is possible to perform cross domain image generation with GAN training in an unsupervised fashion i.e. without explicitly forming the input-output pairs. 

Lastly, our analytical results for speckle intensity autocorrelation shows that the speckles formed in the sensor plane for for a specific object class (say structure $5$ in a USAF test target) in the input plane are unique to that particular object class. Based on this we deduce that image pair dataset formation for training a cGAN is quite effective for outputting a reliable trained model owing to this unique characteristic property of speckles.   

Future work will involve studying strongly-scattering dynamic systems using cGANs and other pathways for real time object retrieval. One other area of interest would be real time PSF modeling and rapid field of view reconstructions for a dynamic system.

In summary, cGANs can compensate for the presence of a scattering layer and enable quantitative feature extraction. cGANs can also be rapidly-trained and thus enables opportunities to compensate for scattering in increasingly complex optical systems.

\section*{Acknowledgments}
Siddharth Rawat thanks UNSW Sydney for scholarship support. Anna Wang is a recipient of the UNSW Scientia Program and Australian Research Council Discovery Early Career Award (DE210100291). We would also like to thank Luke Marshall and Christopher Lee for 3D printing the optical hardware for experiments.


\bibliography{ms}

\end{document}